\documentclass[prd,eqsecnum,showpacs,aps,nofootinbib]{revtex4}
\usepackage{latexsym}
\usepackage{graphicx}
\usepackage{color}
\usepackage{dcolumn}% Align table columns on decimal point
\usepackage{bm}% bold math
\def\beq{\begin{equation}}
\def\eeq{\end{equation}}
\def\bea{\begin{eqnarray}}
\def\eea{\end{eqnarray}}
\def\ds{\displaystyle}
\def\pa{\partial}
\def\mn{_{\mu\nu}}

\begin{document}
%\draft
\title{Cosmic Microwave Background Anisotropy from 
Nonlinear Structures in Accelerating Universes}   
\author{Nobuyuki Sakai }
\email{nsakai@e.yamagata-u.ac.jp}
\affiliation{Department of Education, Yamagata University, 
Yamagata 990-8560, Japan}
\author{Kaiki Taro Inoue}
\email{kinoue@phys.kindai.ac.jp}
\affiliation{Department of Science and Engineering, Kinki University,
Higashi-Osaka 577-8502, Japan}  
\date{\today}

\begin{abstract}
We study the cosmic microwave background (CMB) anisotropy due to
spherically symmetric
nonlinear structures in flat universes with dust and a cosmological constant. 
By modeling a time-evolving spherical compensated void/lump by Lemaitre-Tolman-Bondi spacetimes,
we numerically solve the null geodesic equations with the Einstein equations.
We find that a nonlinear void redshifts the CMB photons 
that pass through it regardless of the distance to it.
In contrast, a nonlinear lump blueshifts (or redshifts) the
CMB photons if it is located near (or sufficiently far from) us.
The present analysis comprehensively covers previous works based
 on a thin-shell approximation and a linear/second order
 perturbation method and the effects 
of shell thickness and full nonlinearity.
Our results indicate that, if quasi-linear and large ($\gtrsim100$Mpc)
 voids/lumps would exist, they could be observed as cold or hot 
spots with temperature variance $\gtrsim 10^{-5}$K in the CMB sky.
\end{abstract}

\maketitle
%\pacs{Valid PACS appear here.
%{\tt$\backslash$\string pacs\{\}} should always be input,
%even if empty.}

%\narrowtext

\section{Introduction}
\label{sec:level1}

Recently, much attention has been paid to 
generation of the cosmic microwave background (CMB) 
anisotropy due to nonlinear evolution of   
the gravitational potential, which is called 
the Rees-Sciama (RS) effect \cite{rs}. 
It has been argued that the RS effect only affects the 
angular power spectrum of the CMB anisotropy
at relatively small angular scales $l\gtrsim 3000$ 
\cite{sato,ms,pane}.
However, recent discoveries of the CMB anomalies such as 
octopole planarity, the alignment between quadrupole
and octopole components \cite{teg1},  
anomalously cold spots on angular scales 
$\sim 10^\circ$ \cite{vie1}, 
and asymmetry in the large-angle power
between opposite hemispheres \cite{eri1} hint that 
the RS effect due to large-scale structures could affect the CMB anisotropy
at large angular scales as well \cite{cs}. 
This possibility is also indicated by a recent report that the density of extragalactic radio sources as projected on the sky is anomalously low in the direction towards the cold spot in the CMB map \cite{rud}.

The signatures of the RS effect due to nonlinear  
voids/lumps in the Friedmann-Robertson-Walker (FRW)
universe without a cosmological constant 
has been extensively studied in the literature \cite{sato,ms,pane}. 
Recently, the RS effect due to a quasi-linear void/lump in the FRW universe with a cosmological 
constant has been studied using a thin-shell approximation \cite{is2}
and a second-order perturbation method \cite{ti}.
In order to check the validity and consistency, 
it is of great importance to extend the analyses to solve the
Einstein equations without relying on these approximations.

In this paper, we study the RS effect due to 
nonlinear structures in flat universes with a cosmological constant 
by solving the Einstein equations, 
which incorporate the fully nonlinear regime. 
Specifically we model a compensated spherical nonlinear void/lump by a family of 
Lemaitre-Tolman-Bondi (LTB) spacetimes, and numerically solve the null 
geodesic equations with the Einstein equations.
In \S 2, we derive the Einstein equations and null geodesic equations for LTB
spacetimes and model a compensating spherical void/lump with a smooth 
mass density profile. In \S 3, we show some numerical results.
\S4 is dedicated to concluding remarks. 

%ch2 ----------------------------------------------------------
\section{Model and Basic Equations}
\subsection{Lemaitre-Tolman-Bondi spacetime}

We consider a spherically symmetric spacetime with dust and a
cosmological constant $\Lambda$, which satisfies Einstein equations,
\beq\label{EinEq}
G\mn+\Lambda g\mn=8\pi G\rho u_{\mu}u_{\nu}
\eeq
where $g\mn,~G\mn,~G,~\rho$, and $u_{\mu}$ are the Riemannian metric tensor, the Einstein tensor, 
the gravitational constant, matter density, and the fluid 4-velocity, respectively.

In spherical coordinates $(t,~r,~\theta,~\phi)$, the general solutions are represented by Lemaitre-Tolman-Bondi (LTB) metric,
\beq\label{LTBmetric}
ds^2=-dt^2+{{R'}^2(t,r)\over1+f(r)}dr^2+R^2(t,r)(d\theta^2+\sin^2\theta d\varphi^2),
\eeq
which satisfies
\beq\label{LTBeq1}
\dot R^2={2Gm(r)\over R}+{\Lambda\over3}R^2+f(r),
\eeq\beq\label{LTBeq2}
\rho={m'(r)\over4\pi R^2R'},
\eeq
where $~'\equiv\pa/\pa r$ and $\dot{~}\equiv\pa/\pa t$. 
The solutions contain two arbitrary functions, $f(r)$ and $m(r)$.
If we we give $'\Lambda,~ \rho(t_i,r)$, and the local Hubble parameter $H(t_i,r)\equiv \dot R(t_i,r)/R(t_i,r)$ at the initial time $t=t_i$, 
$m(r)$ and $f(r)$ are determined by (\ref{LTBeq1}) and (\ref{LTBeq2}). 
The radial coordinate $r$ has a gauge degree of freedom, $r\rightarrow r'=[$any function of $r]$; 
here we define $r$ as the areal radius at the initial time: $R(t_i,r)=r$.

Once $m(r)$ and $f(r)$ are determined, the evolution of $R$ is given by (\ref{LTBeq1}) numerically.
Differentiating (\ref{LTBeq1}) with respect to $r$, we obtain
\beq\label{Yrt}
\dot R'={1\over2\dot R}\left(\frac{2Gm'}{R}-\frac{2 G m}{R^2}R'
+f'+\frac23\Lambda RR'\right),
\eeq
which is the evolution equation of $R'$.
Although $R'$ can be calculated also by the finite difference of $R$ with respect with $r$,
the integration of ('\ref{Yrt}) with respect to $t$ gives better precision for $R'$.

\subsection{Modeling a Void/Lump}

Our model is composed of three regions: the outer flat 
FRW spacetime ($V_+$), the inner negatively/positively 
curved FRW spacetime ($V_-$), and the intermediate shell region ($V_s$). 
In $V_{\pm}$, the field equations (\ref{LTBeq1}) and (\ref{LTBeq2}) 
reduce to the Friedmann equations,
\bea\label{Feq1}
H_+^2={8\pi G\rho_+\over3}+{\Lambda\over3},
&&\rho_+\propto {1\over a_+^3},\\
H_-^2={8\pi G\rho_-\over3}+{\Lambda\over3}+{C^2\over a_-^2},
&&\rho_-\propto {1\over a_-^3},
\label{Feq2}\eea
where $C\equiv\sqrt{f(r_-)}/r_-$ is a constant.
Here $r=r_{\pm}$ denotes the boundary between $V_{\pm}$ and $V_s$ whereas $a_{\pm},~H_{\pm}$, and $\rho_{\pm}$ simply mean the quantities in $V_{\pm}$.

The shell $V_s$ is constructed by the LTB spacetime in such a way
that $m(r)$ and $f(r)$ are continuous through $V_{\pm}$. 
At the initial time $t=t_i$, we assume that $\rho_-\approx\rho_+$, 
$H(t_i,r)$=const., and the matter density profile is given by    
\beq\label{rhoi}
\rho(r)=
\left\{\begin{array}{lcl}
\rho_-
&{\rm for} &r\le r_-,\\
\ds{\rho_c-\rho_-\over16}(3X_-^5-10X_-^3+15X_-+8)+\rho_-
&{\rm for} &r_-\le r\le r_c,\\
\ds{\rho_+-\rho_c\over16}(3X_+^5-10X_+^3+15X_++8)+\rho_c
&{\rm for}& r_c\le r\le r_+,\\
\rho_+
&{\rm for} &r\ge r_+,
\end{array}\right.
\eeq
\beq
{\rm with} ~~
r_c\equiv{r_++r_-\over2},~~~ w\equiv{r_+-r_-\over2},~~~
X_{\pm}\equiv{r-r_c\mp w/2\over w/2},
\eeq
Among the parameters above,  $\rho_c\equiv\rho(r_c)$ cannot be fixed in advance. It is determined as an eigenvalue of Einstein equations (\ref{LTBeq1}) and (\ref{LTBeq2}) with the boundary condition $m(r_+)=4\pi G\rho_+ (a_+r_+)^3/3$ and $f(r_+)=0$.
We define the times $t_1,~t_2,~t_3,$ and $t_4$ as follows: the photon passes $r=r_+$ from $V_+$ to $V_s$ at $t=t_1$, passes $r=r_-$ from $V_s$ to $V_-$ at $t=t_2$, passes the other side of $r=r_-$ from $V_-$ to $V_s$ at $t=t_3$ and passes the other side of $r=r_+$ from $V_s$ to $V_+$ at $t=t_4$.
Examples of initial and evolved configurations of $\rho(t,r)$ are shown in Fig.\ 1.

\begin{figure}
\begin{center}
\includegraphics[width=76mm]{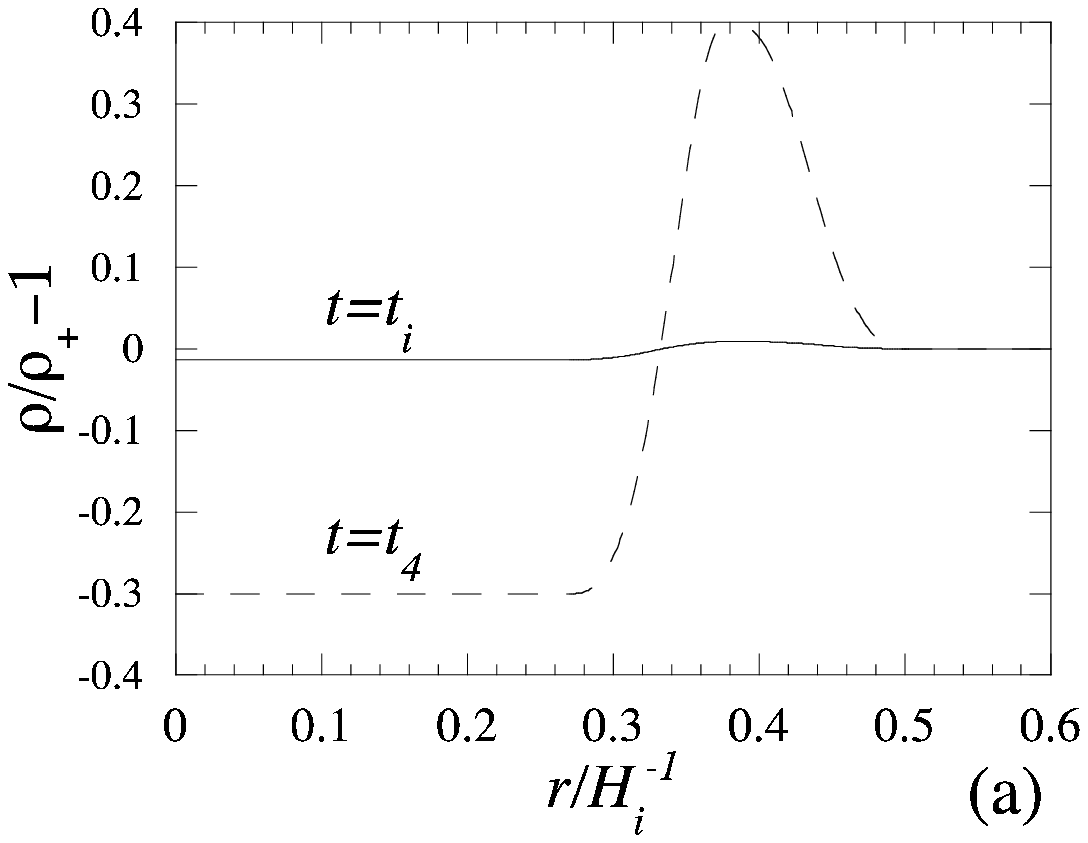}
\hspace*{5mm}
\includegraphics[width=76mm]{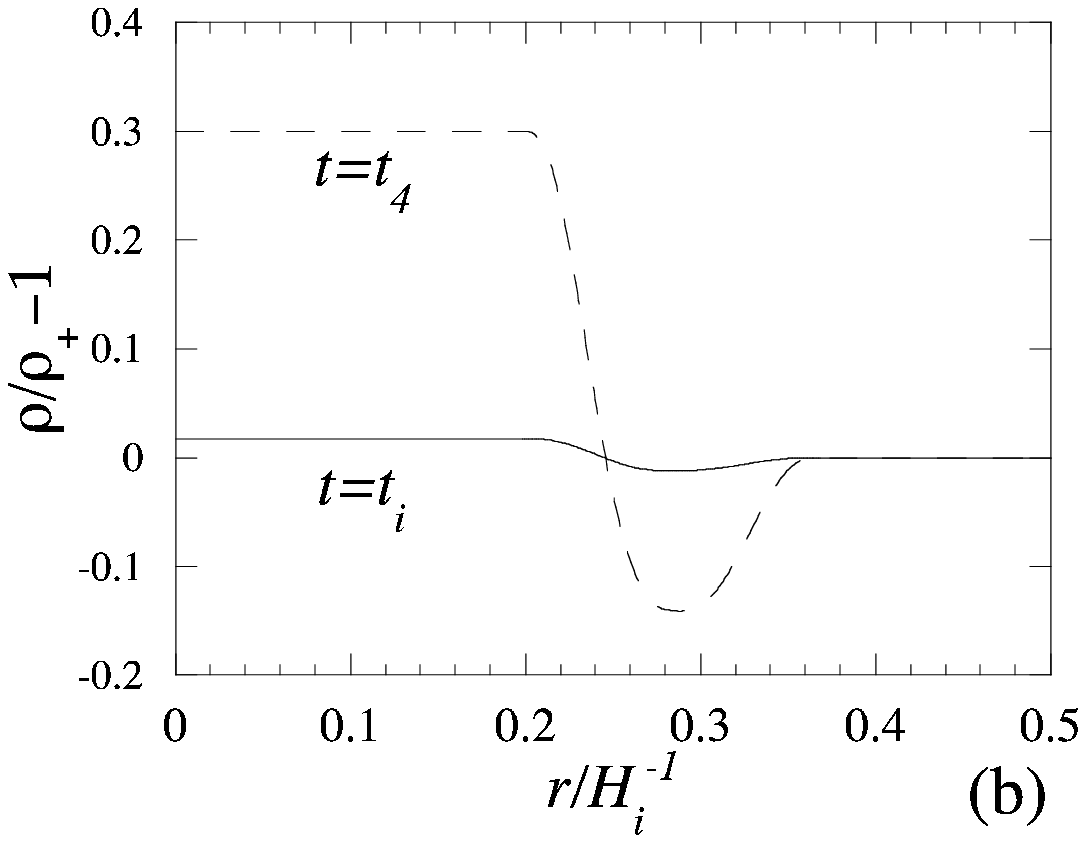}
\caption{\label{f1} Examples of initial and evolved profiles of $\rho(t,r)$.
(a) and (b) correspond a void ($\delta<0$) and a lump ($\delta>0$), respectively.}
\end{center}
\end{figure}

Out model parameters are the background density parameter, the density contrast, the physical radius in 
unit of the Hubble radius, and the width of the shell in unit of the comoving radius of the void/lump,    
\beq
\Omega_4\equiv{8\pi G\rho_+(t_4)\over H_+^2(t_4)},~~
\delta_4\equiv{\rho_-(t_4)\over\rho_+(t_4)}-1,~~
R(t_4,r_c)H_+(t_4),~~ w/r_c,
\label{eq:para}
\eeq
at the exit time $t_4$ of the photon.
The initial parameters $\Omega_i,~\delta_i$, and $r_c$ 
are obtained by iterative integration of the field equation
(\ref{LTBeq1}) with (\ref{eq:para}).

\subsection{Temperature anisotropy}

The wave 4-vector $k^{\mu}$ of a photon satisfies the null
geodesic equations, 
\beq\label{ng1}
k^{\mu}={dx^{\mu}\over d\lambda},~~~ k^{\mu}k_{\mu}=0,
\eeq\beq\label{ng2}
{dk^{\mu}\over d\lambda}+\Gamma^{\mu}_{\nu\sigma}k^{\nu}k^{\sigma}=0,
\eeq
where $\lambda$ is an affine parameter.
In what follows, we only consider a CMB photon which passes the void/lump 
center, $r=0$. Then the geodesic equations (\ref{ng1}) and (\ref{ng2}) with the metric (\ref{LTBmetric}) yield
\beq\label{ng3}
{dt\over d\lambda}=k^t,~~~
{dr\over d\lambda}=k^r,~~~
k^{\theta}=k^{\varphi}=0,
\eeq\beq\label{ng4}
k^r=\epsilon{\sqrt{1+f}\over R'}k^t,~~~
\epsilon\equiv{\rm sign}\left({dr\over dt}\right),
\eeq\beq\label{ng5}
{dk^t\over d\lambda}=-{\dot{g_{rr}}\over2}(k^r)^2,~~~
{d\over d\lambda}(g_{rr}k^r)={{g_{rr}}'\over2}(k^r)^2,~~~
g_{rr}\equiv{(R')^2\over1+f}
\eeq
For the period $t_2<t<t_3$ in $V_-$, the evolution of $k^t_-$ and the
crossing time are given by
\beq
k^t_-\propto{1\over a_-},~~~
\int^{t_3}_{t_2}{dt\over a_-}={2\over C}{\rm arcsinh}(Cr_-).
\eeq

For the periods $t_1<t<t_2$ and $t_3<t<t_4$ in $V_s$ we numerically 
solve the geodesic equations and the field equations simultaneously.
First, we discretize the radial coordinate into $N$ elements, 
\beq
r_i=r_-+(i-1)\Delta r, ~~~ i=1,...,N,~~~ \Delta r={r_+-r_-\over N-1},
\eeq
and $R(t,r)$ into $R_i(t)\equiv R(t,r_i)$.

Next, we rewrite the geodesic equations (\ref{ng3}) - (\ref{ng5}) and the field equations (\ref{LTBeq1}) and (\ref{Yrt}) as differential equations of $r$,
\beq\label{ng6}
{dt\over dr}={\epsilon R'\over\sqrt{1+f}},
\eeq\beq\label{ng7}
{dk^t\over dr}=-{\epsilon\dot R'\over\sqrt{1+f}}k^t,
\eeq\beq\label{ng8}
{d\over dr}(g_{rr}k^r)={{g_{rr}}'\over2}k^r,
\eeq\beq\label{Yi}
{dR_i\over dr}=\dot R_i\left({dt\over dr}\right),
\eeq\beq\label{Yir}
{dR_i'\over dr}=\dot R_i'\left({dt\over dr}\right),
\eeq
where $\dot R$, $\dot R'$  and $\ds\left({dt\over dr}\right)$ are given
by (\ref{LTBeq1}), (\ref{Yrt}) and (\ref{ng6}), respectively. 

Finally, we carry out numerical integration 
of (\ref{ng6}), (\ref{ng7}), (\ref{Yi}), and
(\ref{Yir}) using the fourth-order Runge-Kutta method to obtain 
the solutions of $t(r),~k^t(r),~R_i(t(r))$, and
$R_i'(t(r))$. To estimate the numerical precision, we 
also numerically solve Eq. (\ref{ng8}) and check how 
the solution satisfies (violates) the null condition (\ref{ng4}). 

The energy of a photon passing through the homogeneous 
background without a void/lump is
\beq
k^t_+\propto{1\over a_+}.
\eeq
Then the temperature fluctuation caused by a void/lump 
can be written as 
\beq
{\Delta T\over T}={k^t\over k^t_+}-1.
\eeq

\begin{figure}
\begin{center}
\includegraphics[width=76mm]{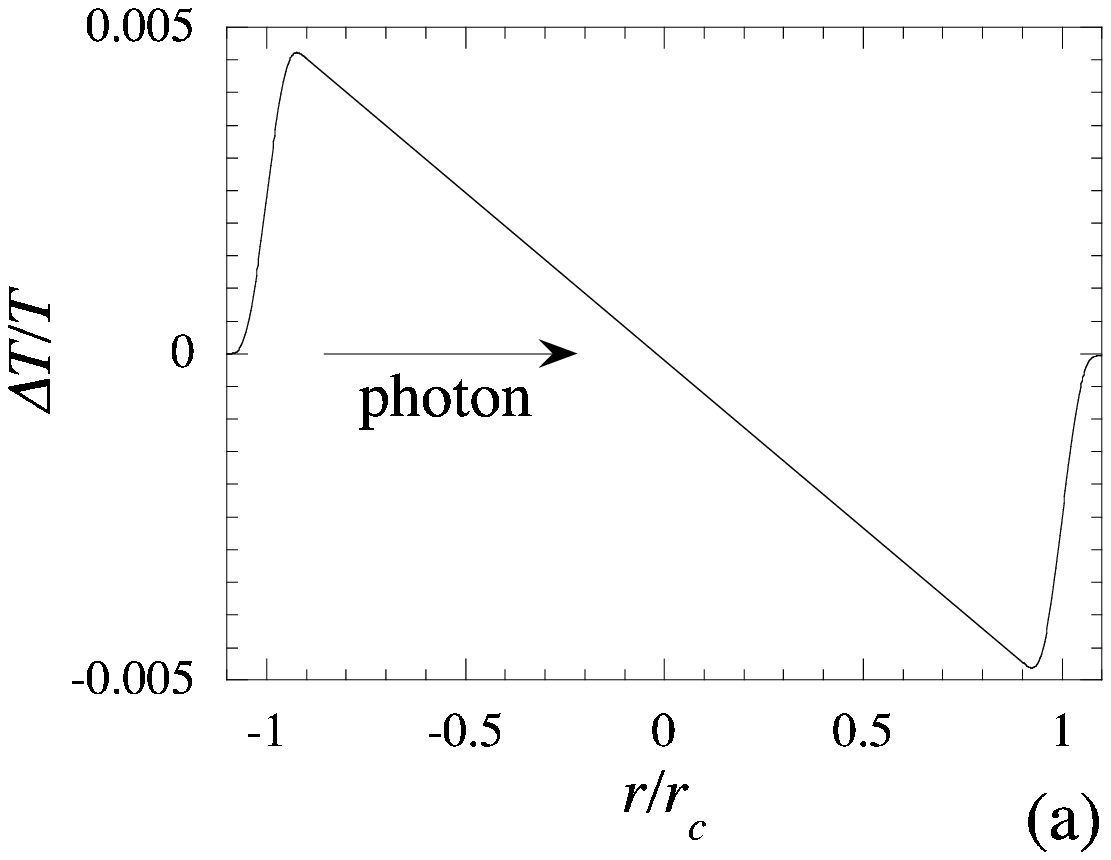}
\hspace*{5mm}
\includegraphics[width=76mm]{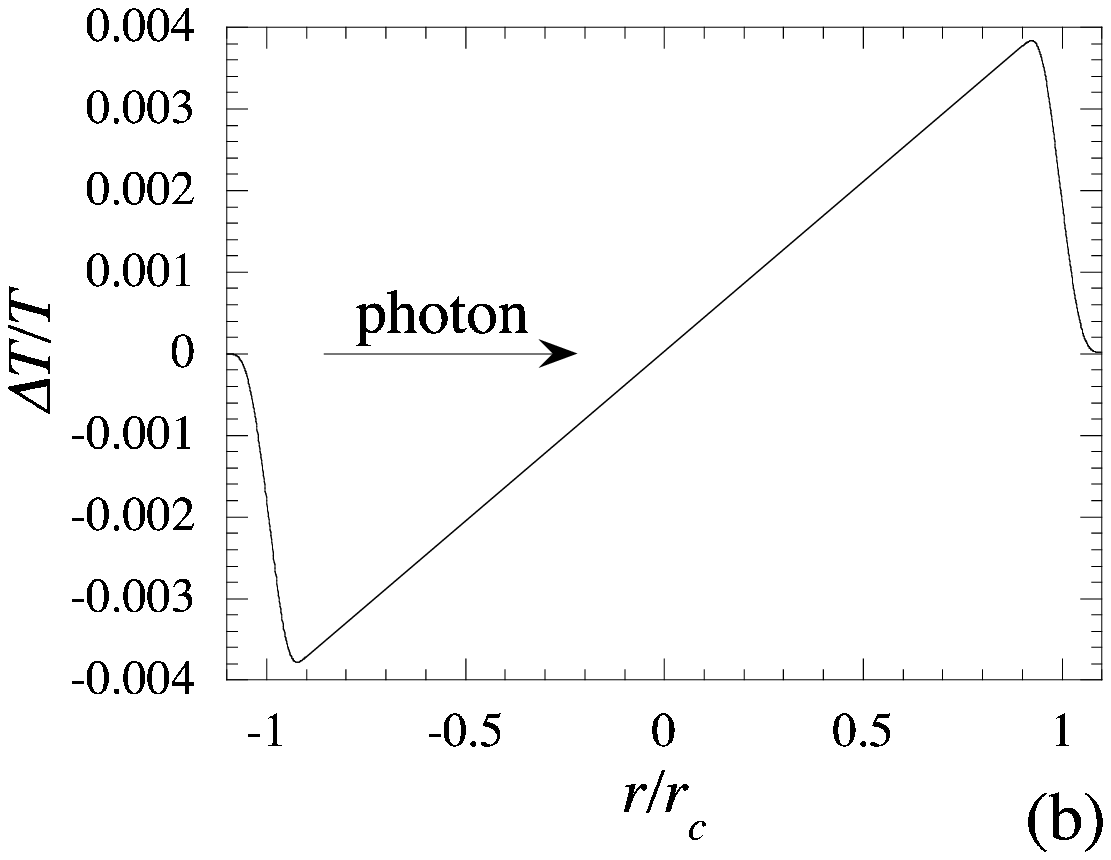}
\vskip -2mm
\caption{\label{f2} Temperature fluctuations of photons passing through
the center of a large void (a) and a large 
lump (b) for comoving observers at each $r=$ constant.
The parameters are  $\delta_4=\mp0.3$, $\Omega_4=0.24$,
 $R_4(r_c)=0.1H_4^{-1}$, and $w/r_c=0.1$. The subscript $4$ denotes 
quantities at the time $t_4$ when a CMB photon exits the edge of a void/lump.
The arrow indicates the traveling direction of a CMB photon.}
\end{center}
\end{figure}

\section{Results}

Figure 2 shows temperature fluctuations of photons passing through
a void/lump for comoving observers at each $r=$ constant.
The amplitude of fluctuations temporarily increases to $|\Delta T/T|\sim10^{-3}$, but it finally
reduces to $\sim10^{-5}$ at the edge of the shell because  
the mass of the void/lump is compensated. 

\begin{figure}
\begin{center}
\includegraphics[width=76mm]{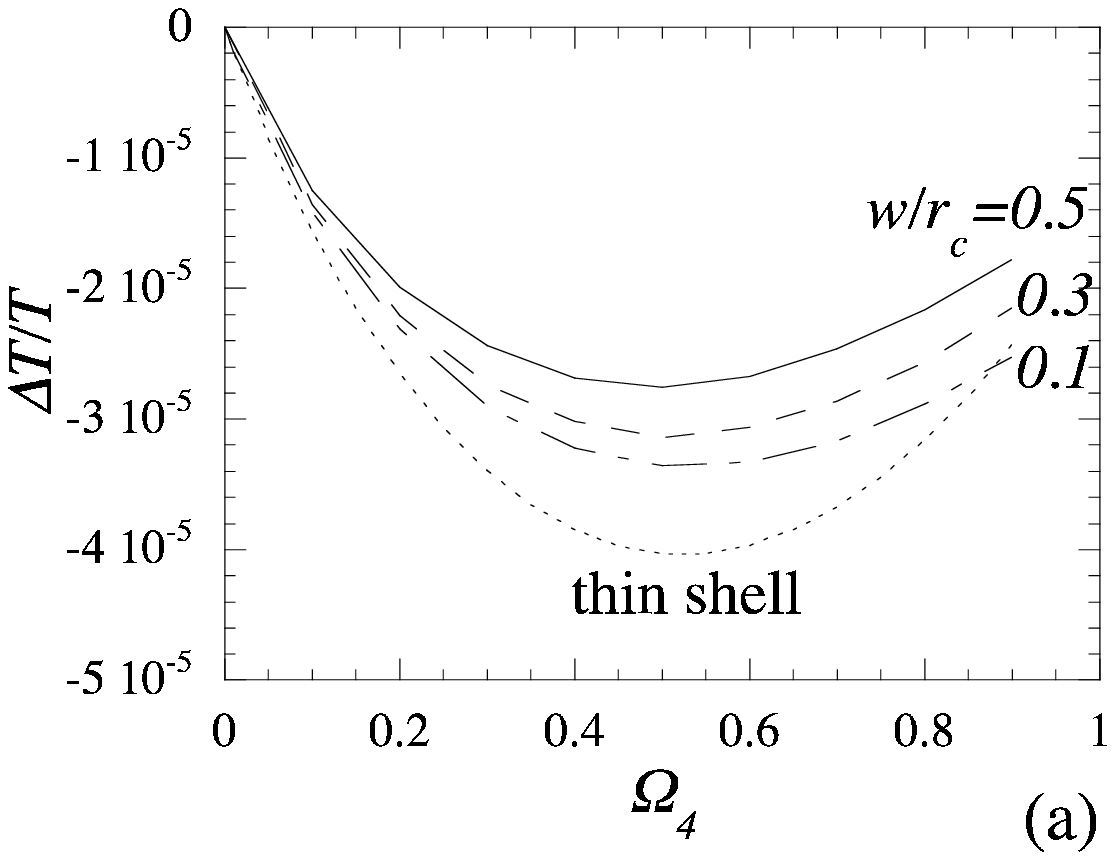}\\
\includegraphics[width=76mm]{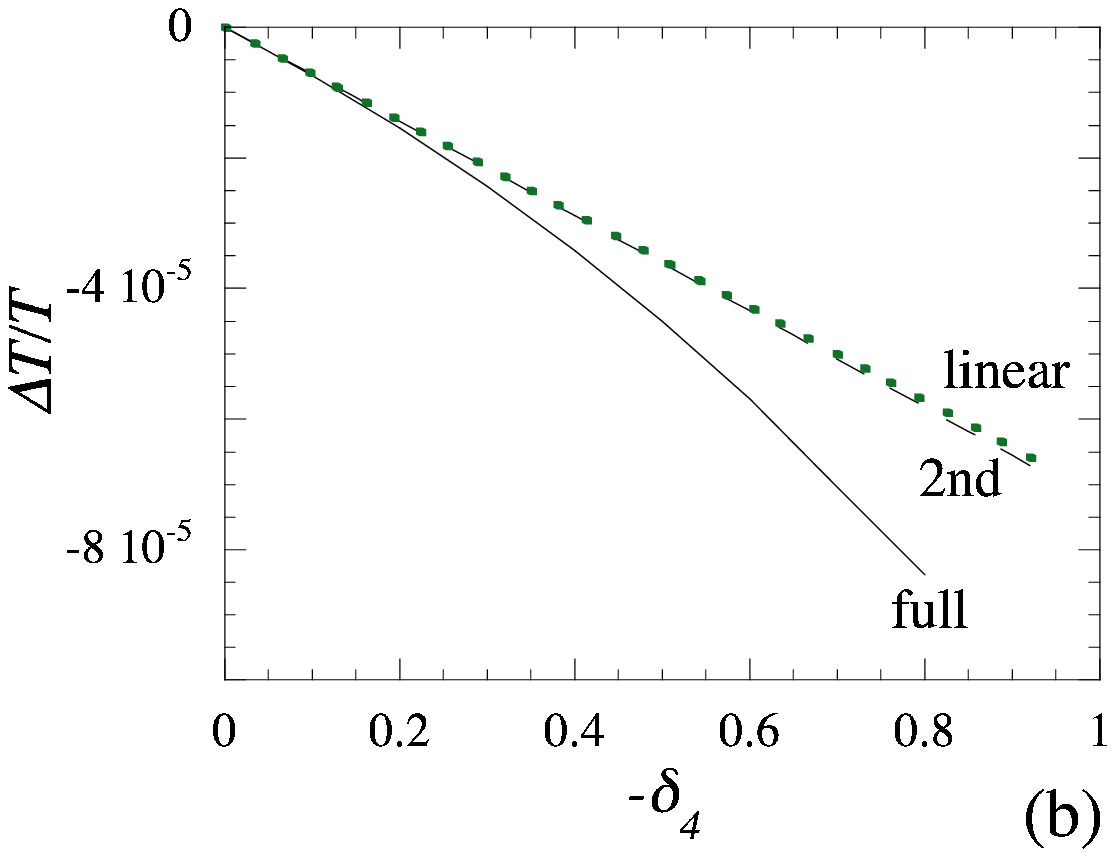}
\hspace*{5mm}
\includegraphics[width=76mm]{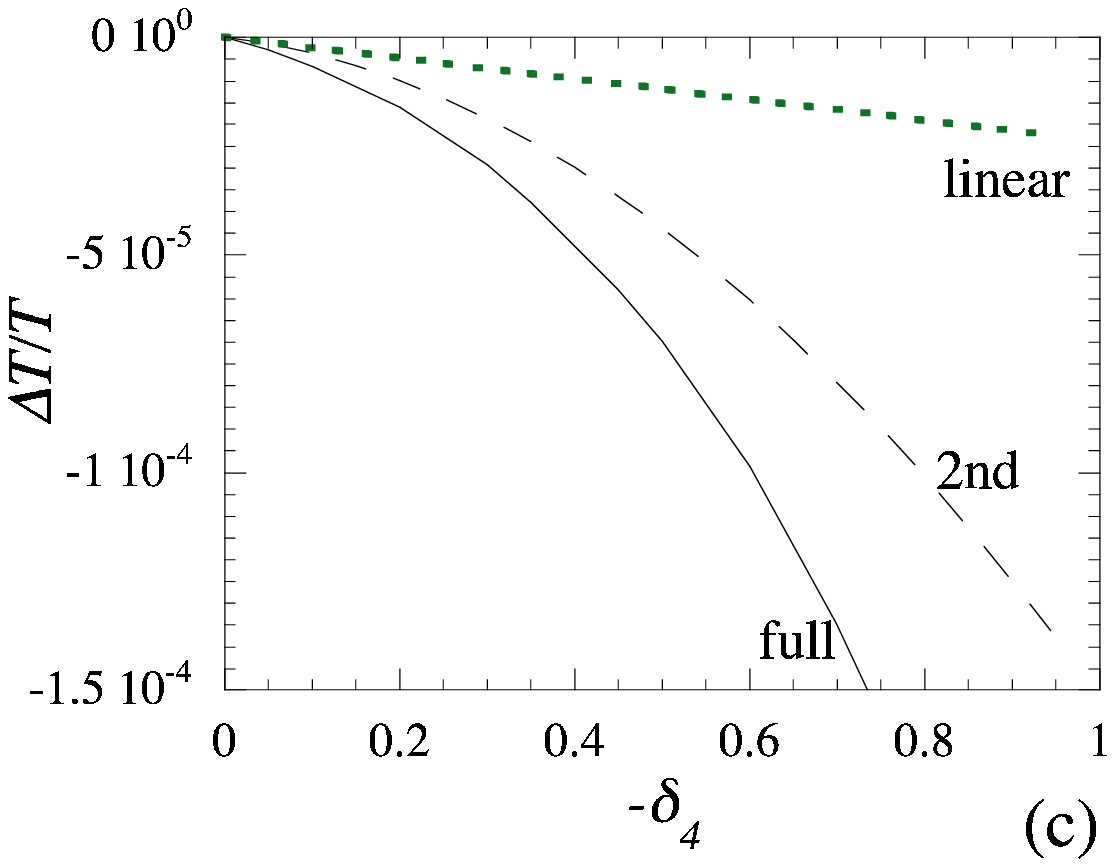}
\vskip -2mm
\caption{\label{f3} Temperature fluctuations of CMB photons passing through
the center of a large void with $R_4(r_c)=0.1H_4^{-1}$ 
for a comoving observer outside the void.
(a) shows $\Delta T/T$ as a function of $\Omega_4$ for $\delta_4=-0.3$.
The dotted line indicated by ``thin shell" shows $\Delta T/T$ for the thin-shell model \cite{is2}.
(b) and (c) show $\Delta T/T$ as a function of $-\delta_4$ for $\Omega_4=0.24$ and for $\Omega_4=0.9$,
respectively; we put $w/r_c=0.3$ for both cases.
The dotted lines and the dashed lines represent the 
values obtained from a linear perturbation analysis, 
and a second order perturbation analysis \cite{ti}, respectively.}
\end{center}
\end{figure}

In what follows, we discuss only the values of $\Delta T/T$
measured by a comoving observer outside a void/lump.
For a void, as Fig.\ 3(a) indicates, the temperature fluctuation $\Delta T/T $ is
always negative regardless of the values of $\Omega_4$.
For a fixed radius, $|\Delta T/T|$ decreases as the width $w/r_c$ of the void shell 
increases. We find that our results are
consistent with those for a thin-shell homogeneous void in the quasi-linear regime \cite{is2}.
To see nonlinear effects, in Fig.\ 3(b)(c) we plot $\Delta T/T$ 
obtained from a linear perturbation analysis, a second order 
perturbation analysis, and our fully nonlinear analysis.
We find that higher-order effects are important and they enhance $|\Delta T/T|$ for a void, particularly for large $\Omega_4$.

\begin{figure}
\begin{center}
\includegraphics[width=76mm]{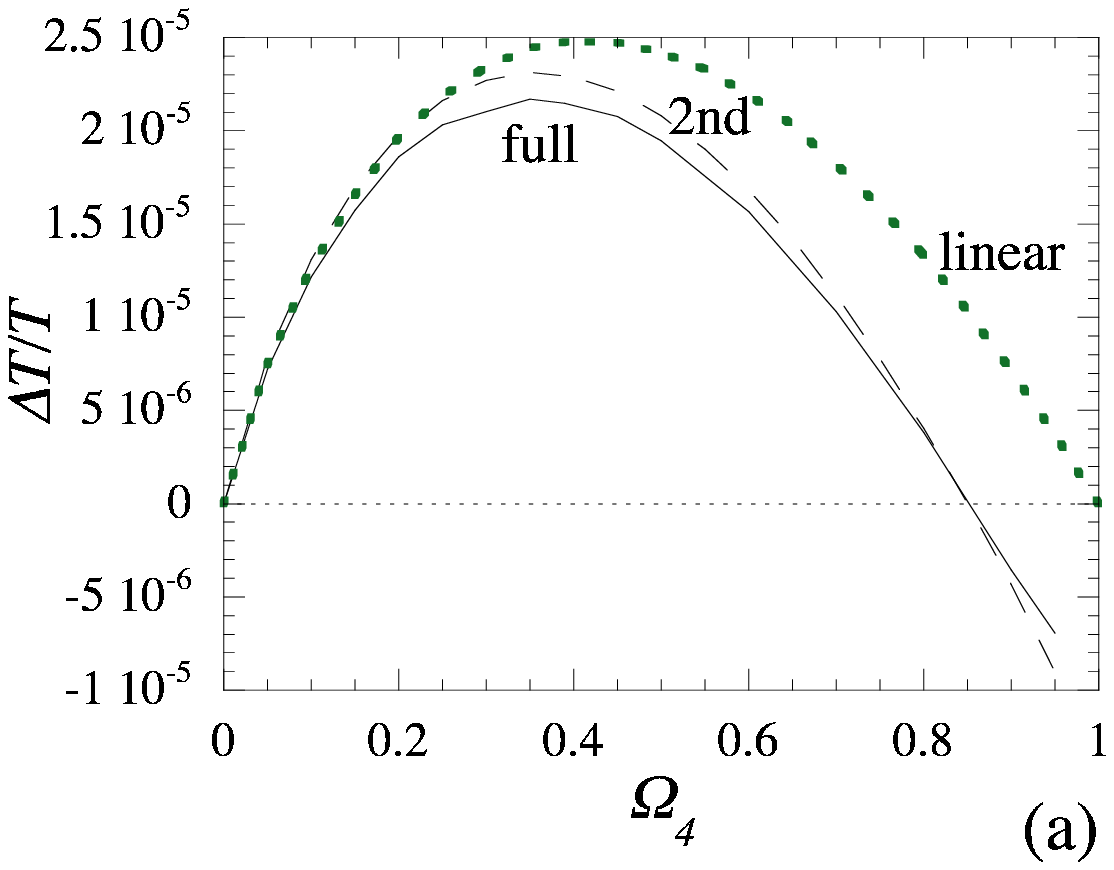}\\
\includegraphics[width=76mm]{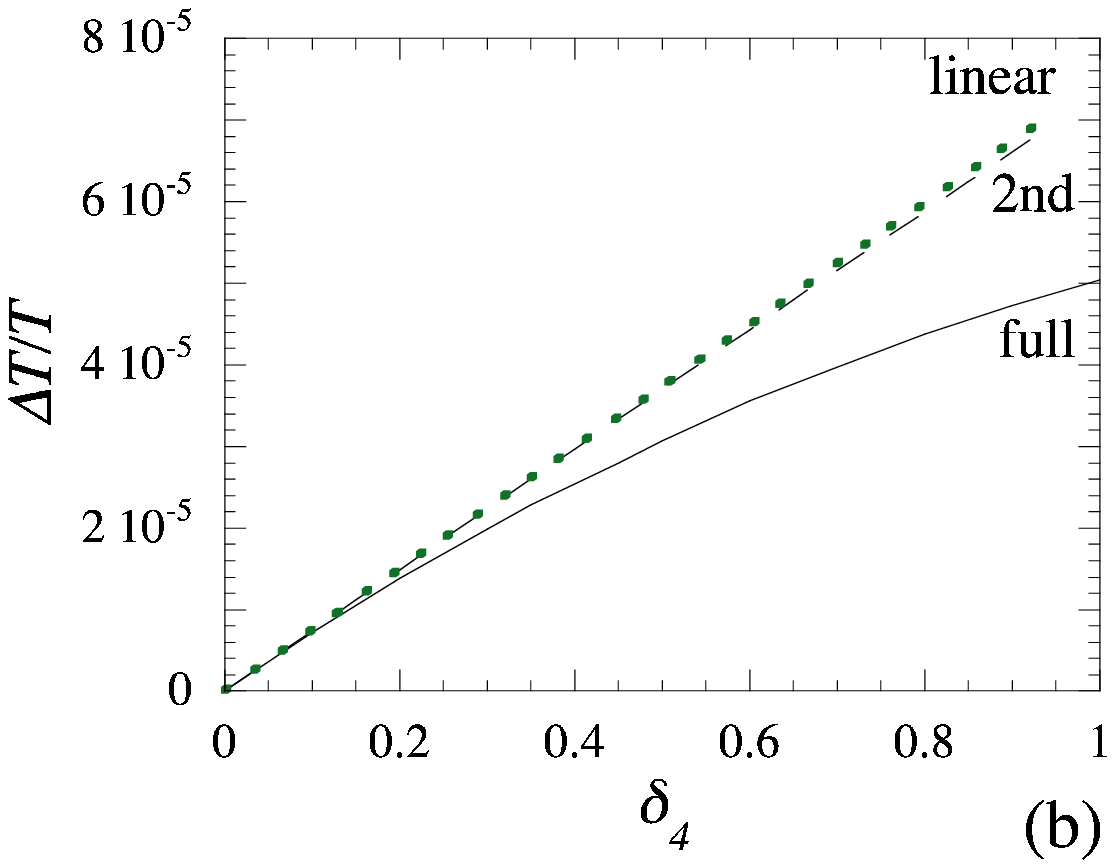}
\hspace*{5mm}
\includegraphics[width=76mm]{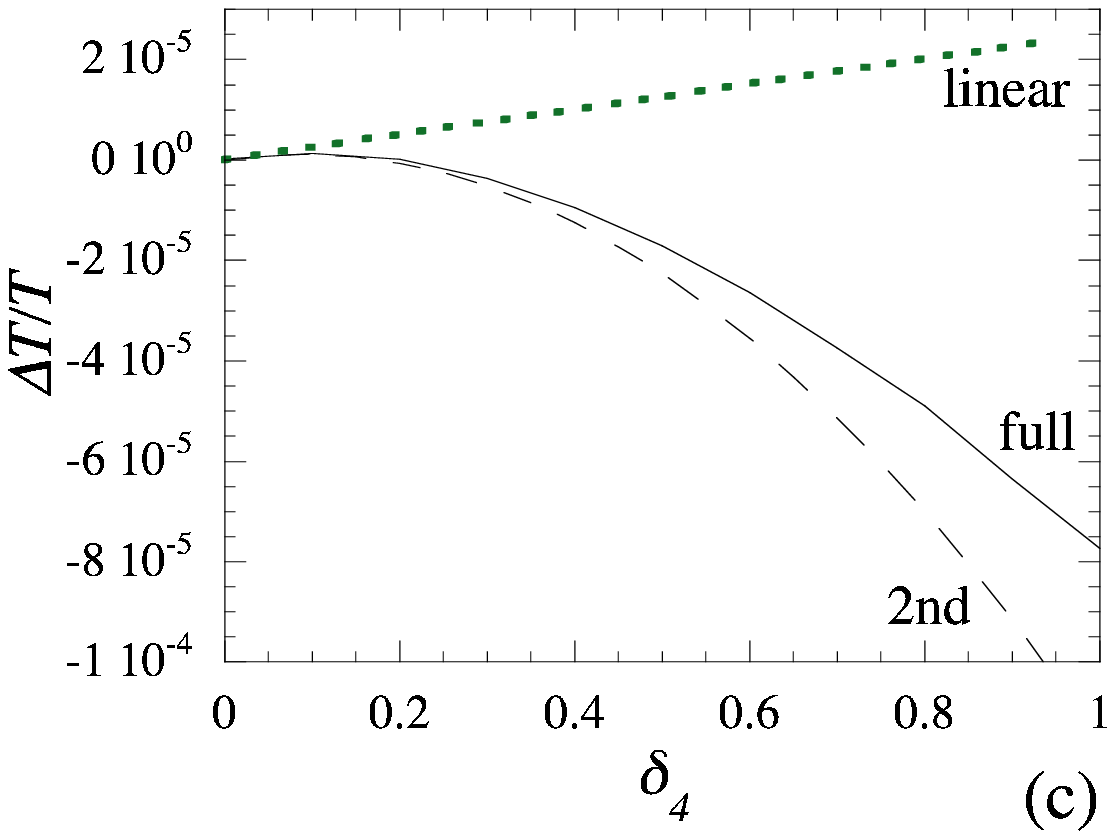}
\vskip -2mm
\caption{\label{f4} Temperature fluctuations of CMB photons 
passing through the center of a large lump with $R_4(r_c)=0.1H_4^{-1}$ and  $w/r_c=0.15$
for a comoving observer outside the lump.
The dotted lines and the dashed lines represent the values 
obtained from a linear perturbation analysis, and a second order 
perturbation analysis \cite{ti}, respectively.
(a) shows $\Delta T/T$ versus $\Omega_4$ for $\delta_4=0.3$.
(b) and (c) show $\Delta T/T$ versus $\delta_4$ for $\Omega_4=0.24$ and for $\Omega_4=0.9$,
respectively.
}
\end{center}
\end{figure}

For a lump, as Fig.\ 4(a) indicates, the temperature fluctuation $\Delta T/T $ is
positive for low background density, but it can be negative for 
high background density.
In other words, lumps at low-$z$ blueshift the CMB
photons, whereas lumps at high-$z$ redshift them. 
This behavior of $\Delta T/T $ in the quasi-linear regime 
can be interpreted as follows.
First, let us consider a perturbative case $|\delta_4|\ll 1$
for which the linear integrated Sachs-Wolfe (ISW) effect \cite{sw}
dominates $\Delta T/T$. 
Then one can show that $\Delta T/T$ vanishes 
for $\Omega_4=0$ (de Sitter) because no matter fluctuation exists,
and for $\Omega_4=1$ (Einstein-de Sitter) because the 
Newtonian gravitational potential freezes in the Einstein-de Sitter universe. 
Therefore, $\Delta T/T$ cannot
be a monotonic function of $\Omega_4$ for $0<\Omega_4<1$. 
In fact, as one can see in Fig.\ 4(a), the ISW contribution
has a peak as a function of $\Omega_4$ for a fixed $\delta_4$. 
Next, let us consider a quasi-linear case $0.1\lesssim|\delta_4|\lesssim 1$.
For small $\Omega_4$, the nonlinear RS effect is not important
because matter fluctuations are small. However, for
large $\Omega_4$, the nonlinear RS effect dominates
the linear ISW effect, which vanishes for $\Omega_4=1$. 
Our numerical analysis shows that the nonlinear RS 
effect reduces the temperature of the CMB photons, which reconfirms the 
previous semi-analytic result for 
lumps in the Einstein-de Sitter universe \cite{ms}.
Thus, one can interpret that 
negative $\Delta T/T$ in the $\Omega_4=1$ background for a void/lump  
(in Fig.\ 3(a)/4(a)) is caused by the nonlinear RS effect alone.  
It should also be noted that our result is consistent with the previous
one obtained from a second order perturbation analysis 
for a void/lump in accelerating universes \cite{ti}.

Fig. 4(b)(c) shows nonlinear effects for a lump:
higher-order effects are still important, but they reduce the 
amplitude of $\Delta T/T$
in contrast to the case for a void in Fig.\ 3(b)(c). 

\section{Concluding Remarks}

We have studied the CMB anisotropy caused by 
spherically symmetric nonlinear structures in flat universes
with dust and cosmological constant.
Specifically, by modeling a time-evolving
spherical compensated void/lump by Lemaitre-Tolman-Bondi spacetimes.
we have solved the null geodesic equations with the Einstein equations numerically.

We have found that a nonlinear void redshifts the 
CMB photons that pass through it regardless of its location.
In contrast, a compensated nonlinear lump blueshifts (or redshifts) the 
CMB photons if it is located near (or sufficiently far from) us.

Our result for the 
temperature anisotropy due to a void is roughly consistent with 
the previous one based on a thin-shell approximation.
We have also shown that $|\Delta T/T|$
decreases as the shell thickness increases for fixed $\delta$.

We have checked that our results 
are also consistent with the ones based on a  
linear/second order perturbation method for small $|\delta|$. 
It turned out that nonlinear (higher-order) effects are important 
even in the quasi-linear regime $|\delta|\gtrsim0.3$.

Our results indicate that, if a quasi-linear ($|\delta|\sim0.3$) and large
size ($R\sim 0.1H^{-1}$) void/lump could exist, 
they would be observed as a cold or hot spot at the level of $\Delta T/T\sim 10^{-5}$ in the CMB sky.
In such a case fully nonlinear and relativistic analysis is necessary.

%%%%%%%%%%%%%%%%%%%%%%%%%%%%%%%%%%%%%%%%%%%%%%%%%%%%%%%%%%%%%%%%%%%%
\acknowledgments{We acknowledge the use of Yukawa Institute Computer 
Facility for implementing numerical computation. 
This work is in part supported by MEXT Grant-in-Aid for Scientific
	Research (C) No.\ 18540248 and for Young Scientists (B)
	No.\ 20740146. }
%%%%%%%%%%%%%%%%%%%%%%%%%%%%%%%%%%%%%%%%%%%%%%%%%%%%%%%%%%%%%%%%%%%%

%ref
%\begin{references}

%*****************************************************

\end{document}